\pdfoutput=1

\documentclass[11pt]{article}

\usepackage[final]{acl}

\usepackage{times}
\usepackage{latexsym}

\usepackage[T1]{fontenc}

\usepackage[utf8]{inputenc}

\usepackage{microtype}

\usepackage{inconsolata}

\usepackage{graphicx}
\usepackage{hyperref}
\usepackage{amsmath,amssymb,amsfonts,graphicx}
\usepackage{algorithmic}
\usepackage{color}
\usepackage{colortbl}
\usepackage{textcomp}
\usepackage{xcolor}
\usepackage{booktabs}
\usepackage{makecell}
\usepackage{url}
\usepackage{multirow}
\usepackage{multicol}
\usepackage{diagbox}
\usepackage{enumitem}
\usepackage{adjustbox}
\usepackage{pifont}
\usepackage{float}
\usepackage{stfloats}
\usepackage[flushleft]{threeparttable}

\title{SpeechFake: A Large-Scale Multilingual Speech Deepfake Dataset Incorporating Cutting-Edge Generation Methods}

\author{
\textbf{Wen Huang\textsuperscript{1,2$\star$}},
\textbf{Yanmei Gu\textsuperscript{2$\star$}},
\textbf{Zhiming Wang\textsuperscript{2}},
\textbf{Huijia Zhu\textsuperscript{2$\dagger$}},
\textbf{Yanmin Qian\textsuperscript{1$\dagger$}}
\thanks{\textsuperscript{$\star$}Equal contribution. \textsuperscript{$\dagger$}Corresponding authors.}
\\ 
\textsuperscript{1}Auditory Cognition and Computational Acoustics Lab\\
MoE Key Lab of Artificial Intelligence, AI Institute\\
School of Computer Science, Shanghai Jiao Tong University, Shanghai, China\\
\textsuperscript{2}Ant Group, Shanghai, China
}

\begin{document}
\maketitle

\begin{abstract}
As speech generation technology advances, the risk of misuse through deepfake audio has become a pressing concern, which underscores the critical need for robust detection systems. However, many existing speech deepfake datasets are limited in scale and diversity, making it challenging to train models that can generalize well to unseen deepfakes. 
To address these gaps, we introduce SpeechFake, a large-scale dataset designed specifically for speech deepfake detection.
SpeechFake includes over 3 million deepfake samples, totaling more than 3,000 hours of audio, generated using 40 different speech synthesis tools. The dataset encompasses a wide range of generation techniques, including text-to-speech, voice conversion, and neural vocoder, incorporating the latest cutting-edge methods. It also provides multilingual support, spanning 46 languages.
In this paper, we offer a detailed overview of the dataset’s creation, composition, and statistics. We also present baseline results by training detection models on SpeechFake, demonstrating strong performance on both its own test sets and various unseen test sets. Additionally, we conduct experiments to rigorously explore how generation methods, language diversity, and speaker variation affect detection performance.
We believe SpeechFake will be a valuable resource for advancing speech deepfake detection and developing more robust models for evolving generation techniques\footnote{Dataset is released at: \url{https://github.com/YMLLG/SpeechFake}}.
\end{abstract}

\begin{table*}[tb]
    \centering
    \caption{Basic statistics of SpeechFake and its comparison with existing speech deepfake datasets. \#utt, \#spk, \#gen represent number of utterances, speakers and generators, respectively. ``-” indicates that the dataset either does not provide information on the number of speakers or generators, or the generator type is unspecified.}
    \label{tab:overall}
    \resizebox{\textwidth}{!}{
    \begin{tabular}{l||c|c|c|c|c|c|c}
    \Xhline{1px} 
    \multirow{2}{*}{\textbf{Dataset}} & \multirow{2}{*}{\textbf{Year}} & \multicolumn{3}{c|}{\textbf{Deepfake Statistics}} & \multirow{2}{*}{\textbf{Generator Types}} & \multirow{2}{*}{\textbf{Languages}} & \multirow{2}{*}{\textbf{Access}} \\ \cline{3-5}
     & & \textbf{\#utt} & \textbf{\#spk} & \textbf{\#gen} & & \\
    \hline\hline 
    ASVspoof2015~\cite{wu2014asvspoof} & 2015 & 246,500 & 106 & 10 & TTS, VC & English & Public \\ 
    FakeOrReal~\cite{reimao2019dataset} & 2019 & 87,285 & 33 & 7 & TTS & English & Public \\
    ASVspoof2019-LA~\cite{nautsch2021asvspoof} & 2019 & 130,378 & 107 & 19 & TTS, VC & English & Public \\ 
    WaveFake~\cite{frank2021wavefake} & 2021 & 117,985 & 2 & 6 & NV & English, Japanese & Public \\ 
    ASVspoof2021-LA~\cite{yamagishi2021asvspoof} & 2021 & 148,148 & 67 & 13 & TTS, VC & English & Public \\ 
    ASVspoof2021-DF~\cite{yamagishi2021asvspoof} & 2021 & 572,616 & 93 & 100+ & TTS, VC & English & Public \\ 
    ADD2022~\cite{yi2022add} & 2022 & 389,419 & - & - & TTS, VC & Chinese & Public \\ 
    CFAD~\cite{ma2024cfad} & 2022 & 231,600 & 279 & 12 & TTS & Chinese & Public \\ 
    In-the-Wild~\cite{muller22does}& 2022 & 11,816 & 58 & - & - & English & Public \\
    ADD2023~\cite{yi2024add} & 2023 & 273,847 & - & - & TTS, VC & Chinese & Public \\  
    HABLA~\cite{tamayoflorez2023habla} & 2023 & 58,000 & 162 & 6 & TTS, VC & Spanish & Public \\ 
    MLAAD~\cite{muller2024mlaad} & 2024 & 82,000 & - & 26 & TTS & 23 Languages & Public \\ 
    CD-ADD~\cite{li2024cross} & 2024 & 117,720 & - & 5 & TTS & Chinese & Public \\ 
    ASVspoof5~\cite{wang2024asvspoof} & 2024 & 1,211,186 & 1,922 & 32 & TTS, VC, AT$^\star$ & English & Public \\ 
    VoiceWukong~\cite{yan2024voicewukong} & 2024 & 413,400 & - & 34 & TTS, VC & English, Chinese & Restricted \\ 
    DFADD~\cite{du2024dfadd} & 2024 & 163,500 & 109 & 5 & TTS & English & Public \\
    CVoiceFake~\cite{li2024safeear} & 2024 & 1,254,893 & - & 6 & NV & 5 Languages & Public \\
    SpoofCeleb~\cite{jung2025spoofceleb} & 2024 & 2,687,292 & 1,251 & 23 & TTS & English & Public \\
    \hline
    SpeechFake-BD & 2025 & 2,003,016 & 541 & 40 & TTS, VC, NV & English, Chinese & \multirow{2}{*}{Public} \\ 
    SpeechFake-MD & 2025 & 1,335,492 & 179 & 6 & TTS, VC &  46 Languages &  \\ 

    \Xhline{1px}
    \end{tabular}}
    \begin{tablenotes}
    \item\scriptsize $^\star$AT: Adversarial Attacks using Malafide~\cite{panariello2023malafide} or Malocopula~\cite{todisco2024malacopula}.
    \end{tablenotes}
\end{table*}

\section{Introduction}
In recent years, speech generation technology has rapidly advanced, with models in text-to-speech and voice conversion systems producing highly natural and high-quality voices~\cite{tan2021survey, triantafyllopoulos2023overview, ju2024naturalspeech}. These systems are increasingly used in virtual assistants, content creation, and language learning, making speech synthesis more accessible and widely adopted. However, as the realism of synthetic voices improves, so does the risk of misuse, especially through speech deepfakes, where synthetic voices are used to impersonate real individuals. Such deepfakes have been employed in fraud~\cite{stupp2019fraudsters}, identity theft~\cite{korshunov2018deepfakes}, and misinformation~\cite{chesney2019deepfakes}, highlighting the significant harm they can cause. Therefore, the growing quality and availability of speech generation systems make the need for robust detection methods more urgent than ever.

A key challenge in developing effective deepfake detection methods is the issue of generalization. Detection models often suffer from substantial performance degradation when confronted with unseen deepfakes~\cite{yamagishi2021asvspoof, muller22does}, which underscores the importance of creating comprehensive datasets to support the development of robust detection systems. However, current datasets for this task come with several limitations. As shown in Table~\ref{tab:overall}, many publicly available datasets are relatively small, and the generation techniques they include are often outdated or limited, making it challenging for models to detect more advanced deepfake technologies. Moreover, most datasets primarily focus on English or Chinese, offering limited representation of other languages.

To address these limitations, we propose SpeechFake, a large-scale dataset designed to significantly improve both the scale and diversity of data available for speech deepfake detection. The dataset contains over 3 million speech deepfakes, amounting to more than 3,000 hours. These deepfakes are generated using 30 publicly available speech generation tools and 10 commercial APIs, covering a comprehensive range of speech generation methods and incorporating cutting-edge techniques capable of producing highly realistic synthetic speech. To support multilingual detection and balance language distribution, SpeechFake is divided into two parts: the Bilingual Dataset (BD), focused on English and Chinese, and the Multilingual Dataset (MD), which spans 46 languages, broadening research opportunities in multilingual environments. Furthermore, unlike most existing datasets that offer only binary labels (real / fake), SpeechFake provides rich metadata, including generation methods, voice id, language, and text transcriptions, which facilitates deeper research into the factors that influence deepfake detection and enables other potential use cases.

In addition, we conduct a comprehensive set of experiments to establish a baseline for SpeechFake and explore key factors that influence deepfake detection performance. First, we evaluate the overall performance across multiple datasets to assess how well models trained on SpeechFake generalize to both seen and unseen data, demonstrating strong performance (Section~\ref{subsec:overall}). Next, we analyze cross-generator performance to examine how different speech generation methods affect detection accuracy (Section~\ref{subsec:generator}). We also investigate cross-lingual performance, exploring how models trained on specific languages perform when exposed to deepfakes in other languages (Section~\ref{subsec:language}). Finally, we assess cross-speaker performance to determine the impact of speaker variability on detection robustness (Section~\ref{subsec:speaker}). These experiments establish a strong baseline for SpeechFake and provide valuable insights into the key aspects that influence speech deepfake detection performance.


\begin{figure*}[tb]
    \centering
    \includegraphics[width=\textwidth]{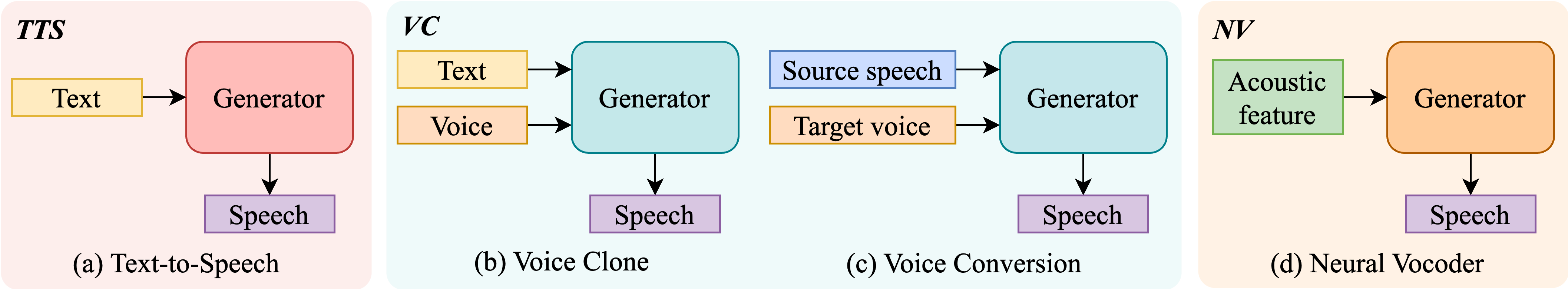}
    \caption{Classification of speech generation methods in SpeechFake based on input modality during inference. (a) TTS: Generate speech from text input. (b)(c) VC: generate speech from text or speech based on target voice. (d) NV: Generate speech from acoustic feature.}
    \label{fig:generator}
\end{figure*}

\section{Related Work}
\label{sec:relate}
\paragraph{Speech Generation}
In prior literature, speech generation (or speech synthesis) is primarily represented by two tasks: Text-to-Speech (TTS) and Voice Conversion (VC). TTS generates speech from text, while VC transforms an existing speech sample to match a target speaker’s voice without altering its linguistic content. These two tasks often share similar model backbones but may differ in task-specific components.

The architecture of TTS has seen significant evolution, starting with CNN/RNN-based models~\cite{oord2016wavenet, wang2017tacotron}, progressing to Transformer-based architectures~\cite{li2019neural, ren2021fastspeech}, and advancing further with generative frameworks such as VAE, GAN, flow, and diffusion models~\cite{prenger2019waveglow, kong2020hifi, kim2021conditional, liu2022diffgan}. Besides, the field has shifted from cascaded acoustic models with separate vocoders~\cite{oord2016wavenet, kong2020hifi} to fully end-to-end systems~\cite{ren2021fastspeech, kim2021conditional}. More recently, the integration of Large Language Models (LLMs) into TTS has enhanced text-to-token generation~\cite{du2024cosyvoice, guo2024fireredtts}.

While traditional TTS systems generated speech in a fixed voice, newer approaches enable multi-speaker synthesis via speaker embeddings~\cite{kim2021conditional, betker2023better} and support few-shot or zero-shot voice cloning, allowing speech generation from minimal target voice samples~\cite{arik2018neural, casanova2022yourtts, wang2023neural, qin2023openvoice}. These advancements have blurred the line between TTS and VC.

VC has undergone a similar architectural transformation. Meanwhile, early methods relied on parallel data and statistical techniques~\cite{godoy2011voice}, whereas modern VC models employ non-parallel training with adversarial and self-supervised learning, significantly improving conversion quality and adaptability~\cite{kaneko2018cyclegan, li2021stargan}.

A key component in many TTS and VC systems is neural vocoders (NV), which generate waveforms from acoustic features (e.g., mel-spectrograms)~\cite{kong2020hifi, lee2023bigvgan}. Traditionally integrated within TTS and VC pipelines, vocoders were not regarded as standalone systems. However, recent studies indicate that vocoded audio also plays a crucial role in deepfake detection~\cite{frank2021wavefake, wang2023spoofed, wang2024can}. 

To better align speech generation with deepfake research, we categorize speech generation methods into three types based on input modality at inference, as shown in Figure~\ref{fig:generator}: TTS, VC (Voice Clone or Voice Conversion), and NV (Neural Vocoder). In this classification, TTS refers to systems that generate speech with seen voices from text alone. VC focuses on generating speech with target voice reference, whether the content comes from text or  speech. Finally, NV generates speech from acoustic features without explicitly altering the original voice.  
By adopting this classification, we encompass a more comprehensive and systematic framework for deepfake speech generation.

\paragraph{Speech Deepfake Datasets}
Several benchmark datasets have been developed for speech deepfake detection. The ASVspoof Challenge series~\cite{wu2014asvspoof, nautsch2021asvspoof, yamagishi2021asvspoof, wang2024asvspoof} has progressively expanded from spoofing attacks on automatic speaker verification (ASV) systems to a broader range of speech deepfakes. Similarly, the Audio Deepfake Detection (ADD) Challenge has released datasets focusing on deepfake detection in Chinese~\cite{yi2022add, yi2024add}.
Other datasets include FoR~\cite{reimao2019dataset}, WaveFake~\cite{frank2021wavefake}, and In-the-Wild~\cite{muller22does}, which collect deepfake speech from various synthesis methods, including open-source tools, neural vocoders, and internet sources. Multilingual resources such as HABLA (Spanish)~\cite{tamayoflorez2023habla}, MLADD (23 languages)~\cite{muller2024mlaad}, and CVoiceFake (5 languages)~\cite{li2024safeear} further extend language coverage.

Meanwhile, recent datasets have gradually integrated advanced speech synthesis techniques. CD-ADD~\cite{li2024cross} targets zero-shot TTS, DFADD~\cite{du2024dfadd} focuses on diffusion-based models, VoiceWukong~\cite{yan2024voicewukong} covers various synthesis methods with perturbation variants, and SpoofCeleb~\cite{jung2025spoofceleb} provides speaker-dependent deepfakes generated from real-world and TTS-based samples.

Despite the availability of several datasets, none offer a comprehensive combination of both large-scale data and diversity in generation methods and languages. Simply merging existing datasets to create a larger benchmark would likely introduce issues such as condition mismatches and increased complexity in model training. SpeechFake addresses these challenges by offering a large-scale, multilingual dataset that incorporates cutting-edge synthesis techniques, providing broader and more robust coverage for deepfake detection research.

\begin{figure*}[tb]
    \centering
    \includegraphics[width=\textwidth]{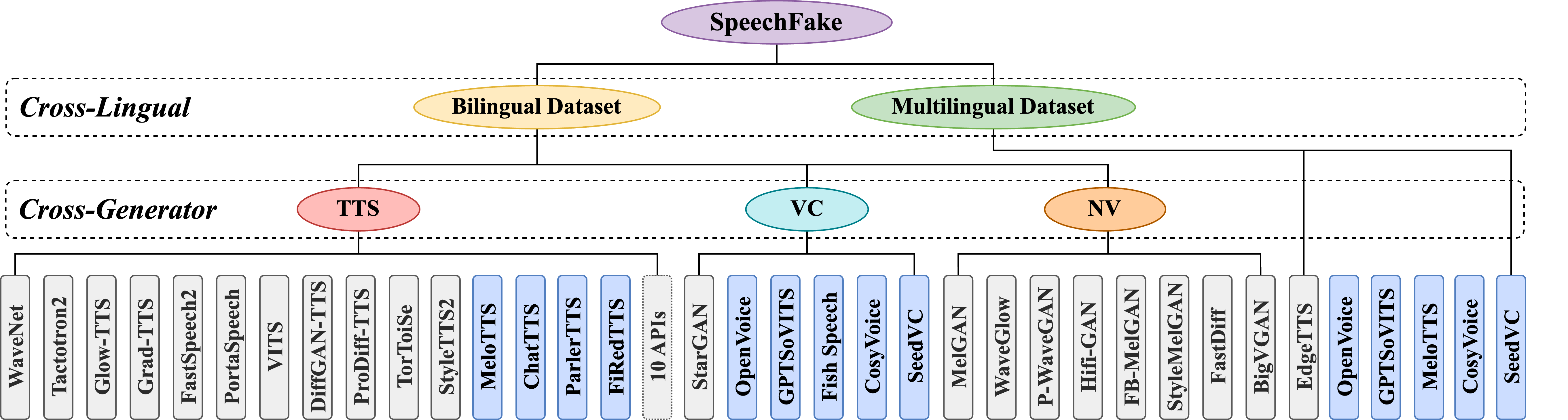}
    \caption{Overview of the SpeechFake dataset. The dataset is divided into two parts: the Bilingual Dataset and the Multilingual Dataset. The Bilingual Dataset is further categorized into three generation methods: TTS, VC, and NV. Methods highlighted in blue represent the latest speech generation methods.}
    \label{fig:dataset}
\end{figure*}

\section{Dataset Collection and Statistics}
\subsection{Data Collection}
\label{sec:collect}

The data collection consists of two parts: real speech, sourced from existing datasets, and fake speech, generated using open-source speech generation methods or commercial APIs. Since most speech generation methods primarily support English or Chinese, we split our dataset into two parts to balance the samples for each language: the Bilingual Dataset (BD), which includes English and Chinese, and the Multilingual Dataset (MD), which covers data from 46 languages. An overview of the dataset composition is shown in Figure~\ref{fig:dataset}.

For BD, real speech data is sourced from four datasets: LibriTTS~\cite{zen2019libritts} and VCTK~\cite{veaux2013voice} for English, and AISHELL1~\cite{bu2017aishell} and AISHELL3~\cite{shi2020aishell} for Chinese. Fake speech is generated using 30 open-source speech generation tools and 10 commercial APIs, as detailed in Table~\ref{tab:list}.
The open-source models span a variety of architectures, including GAN-based models~\cite{kumar2019melgan, kong2020hifi}, Diffusion models~\cite{liu2022diffgan, huang2022prodiff}, Sequence-to-Sequence models~\cite{oord2016wavenet, ren2021fastspeech}, and Flow or VAE models~\cite{prenger2019waveglow, kim2021conditional}. Besides, we include the latest speech generation techniques (highlighted in blue in Figure~\ref{fig:dataset}), all of which were released in the past year and represent cutting-edge advancements in speech synthesis.

\begin{figure*}[tb]
    \centering
    \includegraphics[width=\textwidth]{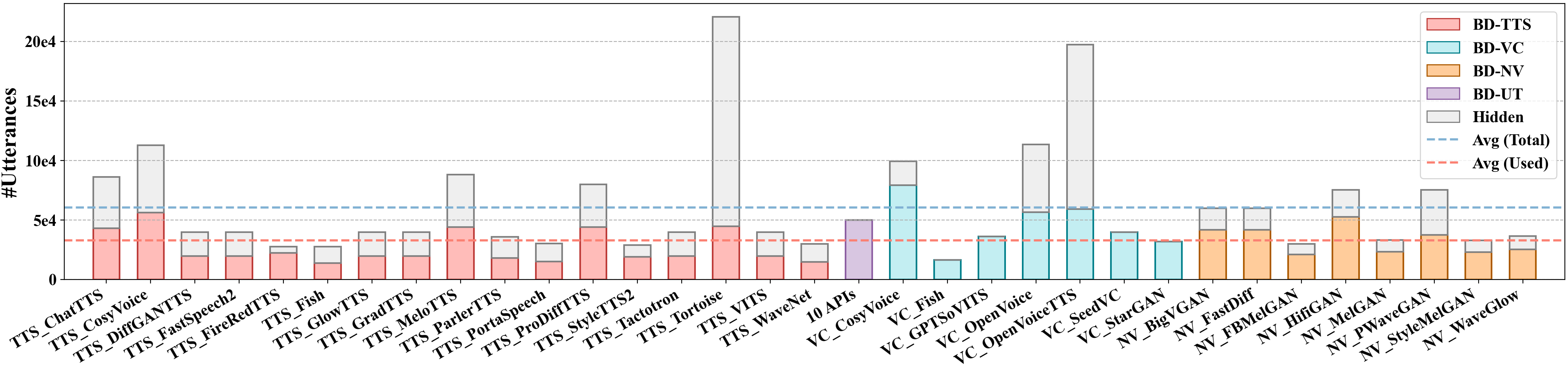}
    \caption{Distribution of speech generation methods in SpeechFake-BD. Some data is hidden in experiment trials to ensure a more balanced distribution across each method and across different test trials.}
    \label{fig:methods}
\end{figure*}

\begin{figure}[tb]
    \centering
    \includegraphics[width=0.48\textwidth]{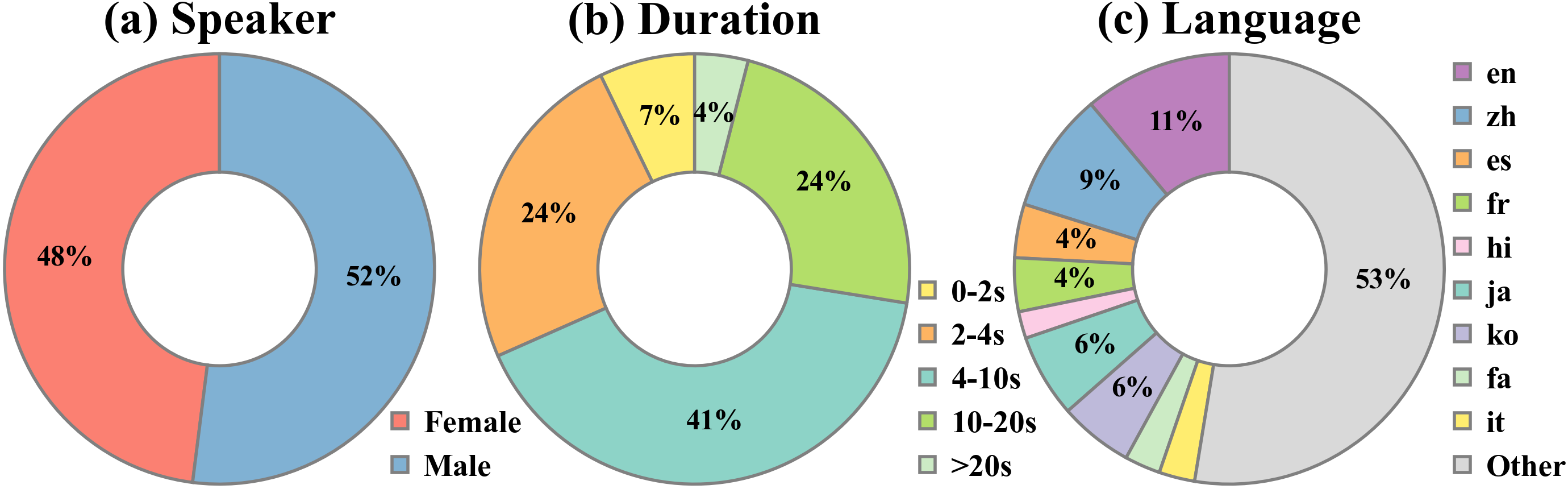}
    \caption{Distribution of speaker gender, language, and duration in SpeechFake. The speaker and duration statistics are based on the entire dataset, while the language distribution is specific to the MD subset.}
    \label{fig:pie}
\end{figure}

For MD, real speech data is sourced from the CommonVoice dataset~\cite{ardila2019common}, which supports multiple languages. Fake speech data is generated using 6 multilingual speech generation tools, as shown in Figure~\ref{fig:dataset}. EdgeTTS\footnote{\url{https://github.com/rany2/edge-tts.git}} supports the widest range of languages, while the other tools cover a subset based on their respective multilingual capabilities.

\paragraph{Data Preparation}
Before generation, we prepare the necessary text or audio inputs for each generator. These inputs are sourced from real datasets, including text transcriptions for TTS systems and audio samples for VC and NV systems.
\vspace{-1mm}
\begin{itemize}[leftmargin=*,itemsep=0pt]
\item \textbf{Text Preprocessing}: For TTS systems, we clean the text inputs by removing special characters, punctuation, and extra spaces. We also ensure that each text sample maintains an appropriate word or character count (e.g., 5–30 words for English) and provides a broad phonemic coverage. The text is then tokenized and formatted to meet the specific requirements of each TTS model, with adjustments made for sentence length or phonetic transcription where necessary.
\item \textbf{Audio Preprocessing}: Audio samples for VC and NV systems are resampled to the required sampling rate and converted into the appropriate formats, such as mel-spectrograms for neural vocoders or raw waveforms for voice conversion models. The silence at the beginning or end of the clips is trimmed.
\end{itemize}

\vspace{-1mm}
\paragraph{Data Generation} 
During the data generation process, the prepared inputs are fed into the respective generators based on the system type.
\vspace{-1mm}
\begin{itemize}[leftmargin=*,itemsep=0pt]
    \item \textbf{For TTS systems}: The prepared text is used to generate speech for each method. If the method supports multiple voices, the text is evenly split among the available voices. An exception is TTS\_Tortoise, for which additional data is generated to support the cross-speaker experiment.
    \item \textbf{For VC systems}: Reference voices are sampled from the real datasets, while the content comes from the selected text or the corresponding speech recordings. The text is generally split equally among the reference voices. For methods supporting style transfer (e.g., CosyVoice, OpenVoice), we include additional data to reflect the transformed styles.
    \item \textbf{For NV systems}: The generated speech is based on the original input audio selected from the real datasets, without any explicit instructions to alter the content or voice.
\end{itemize}

\vspace{-1mm}
\paragraph{Data Post-processing}
Once the speech is generated, we perform several post-processing steps to ensure that the data meets the required quality standards and is suitable for downstream tasks:
\vspace{-1mm}
\begin{itemize}[leftmargin=*,itemsep=0pt]
\item \textbf{Quality Filtering}: We apply voice activity detection (VAD) to filter out speech segments shorter than 0.5 seconds. Additionally, a selective human review is conducted to discard generated speech with noticeable distortions, excessive noise, or unnatural artifacts. Approximately 1\% of samples from each method are randomly selected to ensure representative coverage of the diversity in generation options (e.g., voices, languages).
\item \textbf{Format Standardization}: The remaining audio clips are standardized to a 16kHz sampling rate, converted to mono, and saved in WAV format to ensure consistency across all samples in the dataset.
\end{itemize}

\subsection{Dataset Statistics}
The dataset partitioning for experiments is outlined in Table~\ref{tab:basic} in the Appendix. To address the imbalance between the substantial amount of fake data and the limited real data, as well as to ensure balanced test trials, we allocated approximately half of the fake data across the train, dev, and test sets (split 6:1:3), reserving the remainder for future experiments.

The main portion of BD (train, dev, and test) includes speech deepfakes generated using open-source speech generation methods. It is divided into two language partitions: BD-EN (English) and BD-CN (Chinese), as well as three generator-based subsets: BD-TTS, BD-VC, and BD-NV. To assess model generalization to unseen methods, we also created a separate unseen test set (BD-UT) using commercial APIs. Figure~\ref{fig:methods} illustrates the distribution of the different generation methods used in BD. TTS methods account for the majority, while VC and NV methods represent a smaller portion. On average, each method generates around 60k utterances, with 30k samples used per method for balanced trials.

MD utilizes the same set of generation methods with multilingual support across its train, dev, and test sets, with the key distinction being language coverage. The dataset spans 46 languages, including 9 primary languages with larger sample volumes: English (en), Chinese (zh), Spanish (es), French (fr), Hindi (hi), Japanese (ja), Korean (ko), Persian (fa), and Italian (it). As shown in Figure~\ref{fig:pie} (c), these 9 languages account for half of the dataset, with English and Chinese being the most prominent, while the remaining 37 languages make up the other half.
The train and dev sets consist exclusively of English and Chinese data, while the test set is divided into 10 subsets: one for each of the 9 primary languages and one combined subset for the remaining 37 languages. For the combined subset, around 5,000 clips are selected per language, with the rest reserved for future research.

Figure~\ref{fig:pie} also illustrates the distribution of speaker gender and audio duration. We ensure a relatively balanced representation of female and male speakers in terms of gender. Regarding audio duration, most clips range from 2.0 to 20.0 seconds, with a smaller number of shorter clips (0–2 seconds) and longer ones (over 20 seconds), providing variability in length.

\section{Experiments and Analysis}
\subsection{Experimental Settings}
To evaluate deepfake detection performance, we use two state-of-the-art models: AASIST~\cite{jung2022aasist} and W2V+AASIST~\cite{tak2022automatic}. AASIST employs a heterogeneous stacking graph attention network with a novel attention mechanism to capture spoofing artifacts across both temporal and spectral domains. W2V+AASIST integrates Wav2Vec2.0 XLSR~\cite{babu2021xlsr} as a frontend feature extractor with AASIST serving as the backend classifier. The training details for each model are provided in Table~\ref{tab:train} in the Appendix. For evaluation, we use the Equal Error Rate (EER) as the metric, following previous work~\cite{yamagishi2021asvspoof, du2024dfadd}.

\begin{table*}[t]
\centering
\caption{Performance evaluation (EER\%) of different models trained on ASVspoof2019 (ASV19) or SpeechFake Bilingual Dataset (BD) across multiple test sets, including  subsets of BD and other publicly available benchmarks. For each model, the best results are \textbf{bold} and the second best are \underline{underlined}.}
\label{tab:baseline}
\resizebox{\textwidth}{!}{
\begin{tabular}{c|c|p{1.2cm}<{\centering}|p{1.2cm}<{\centering}|p{1.2cm}<{\centering}|p{1.2cm}<{\centering}|p{1.2cm}<{\centering}|p{1.2cm}<{\centering}|p{1.2cm}<{\centering}|p{1.2cm}<{\centering}|p{1.2cm}<{\centering}|p{1.2cm}<{\centering}}
\Xhline{1px} 
\multirow{2}{*}{\textbf{Train Data}} & \multirow{2}{*}{\textbf{Model}} & \multicolumn{3}{c|}{\textbf{Test Data (SpeechFake)}}  & \multicolumn{7}{c}{\textbf{Test Data (Others)}} \\ \cline{3-12}
 & & BD & BD-EN & BD-CN & ASV19 & FOR & WF & CFAD & ITW & CDADD & ASV24 \\ \hline\hline 
\rowcolor{gray!15}ASV19 & \multirow{4}{*}{AASIST} & 39.36 & 41.05 & 39.07 & 1.88 & 36.08 & 21.17 & 43.95 & 45.27 & 49.53 & 41.89 \\ 
BD & & \textbf{3.48} & \textbf{3.98} & \textbf{2.68} & \underline{23.62} & \textbf{23.35} & \textbf{4.30} & \underline{34.32} & \underline{7.53} & \textbf{22.52} & \underline{35.02}  \\
BD-EN & & \underline{9.02} & \underline{6.17} & 12.00 & 30.65 & 28.99 & 8.54 & 43.39 & \textbf{6.96} & \underline{23.24} & 40.82 \\
BD-CN & & 16.58 & 24.59 & \underline{5.43} & \textbf{16.56} & \underline{25.48} & \underline{5.88} & \textbf{32.34} & 8.54 & 39.75 & \textbf{34.39} \\ 
\hline
\rowcolor{gray!15}ASV19 & \multirow{4}{*}{W2V+AASIST}  &  23.78 & 20.15 & 24.93 & 0.89 & 6.18 & 3.48 & 20.53 & 10.07 & 8.55 & 1.41 \\ 
BD & & \textbf{3.54} & \textbf{3.55} & \textbf{2.83} & \underline{2.91} & \underline{6.00} & \textbf{0.58} & \underline{12.39} & \textbf{2.01} & \textbf{2.42} & \textbf{0.71} \\ 
BD-EN  & & \underline{8.65} & \underline{4.58} & 10.44 & 5.28 & 8.33 & 0.96 & 21.42 & \underline{2.62} & \underline{3.54} & \underline{0.71} \\ 
BD-CN  & & 8.99 & 11.40 & \underline{4.51} & \textbf{0.99} & \textbf{4.88} & \underline{0.64} & \textbf{11.72} & 3.34 & 7.16 & 1.17 \\ 
\Xhline{1px} 
\end{tabular}
}
\end{table*}

\subsection{Overall Performance}
\label{subsec:overall}
We first establish baseline results to demonstrate the overall performance on the Bilingual Dataset. For training, we include the ASVspoof2019-LA training set (ASV19), a widely used benchmark in speech deepfake detection research, alongside three partitions of the BD training set (BD, BD-EN, BD-CN).
The evaluation is conducted on multiple test sets: the BD testing sets (BD, BD-EN, BD-CN), and some additional commonly used benchmarks, spanning a range of datasets from older to newer: ASVspoof2019-LA eval set (ASV19), FakeOrReal (FOR), WaveFake (WF), In-the-Wild (ITW), CDADD, ASVspoof5 (ASV24). Details of the test settings are provided in Appendix~\ref{app:exp}.

From Table~\ref{tab:baseline}, we observe that when models are trained on ASV19, they perform well on its own evaluation set but experience significant performance degradation on other test sets, particularly on BD, where most of the generation methods are unseen during training.
In contrast, training on BD leads to significant accuracy improvements. While training on the English (BD-EN) or Chinese (BD-CN) subsets yields good performance on their respective test sets, it results in poorer performance on the complementary sets. This may be attributed to the differences in the generation methods or languages included in each partition. Using the full BD training set delivers the best overall results, enhancing accuracy across all BD test subsets compared to training on a single language subset. 

When tested on external datasets, models trained on BD consistently outperform those trained on ASV19, except on the ASV19 test set, which is in-domain for the ASV19-trained models. The improvements are particularly significant for test sets such as WF, ITW, and CDADD, where models trained on BD show 50\%-80\% better performance compared to those trained on ASV19.  
Notably, the BD-EN and BD-CN subsets show different performance patterns across test sets. BD-EN performs better on English datasets such as ITW and CDADD, while BD-CN tends to perform better on Chinese datasets like CFAD. However, BD-CN also outperforms both BD-EN and BD on some English test sets, such as ASV19 and FOR. This indicates that language is not the sole factor influencing performance on these unseen test settings. Other factors, such as generation methods and recording conditions, likely contribute as well. Hence, to accurately evaluate the impact of factors like language, other variables should be controlled and kept as consistent as possible across experiments.

\subsection{Cross-Generator Performance}
\label{subsec:generator}
To evaluate the impact of generators on detection performance, we conduct cross-evaluations using three categories of generators in BD: TTS, VC, and NV. The results are presented in Table~\ref{tab:generator}.

\begin{table}[htb]
\centering
\caption{Performance evaluation (EER\%) of different generator types used as training sets across various test sets. For each model, the best results are \textbf{bolded}, and the second-best results are \underline{underlined}.}
\label{tab:generator}
\resizebox{0.48\textwidth}{!}{
\begin{tabular}{c|p{1.3cm}<{\centering}|c|c|c|c|c}
\Xhline{1px} 
\textbf{Train} & \multirow{2}{*}{\textbf{Model}} & \multicolumn{5}{c}{\textbf{Test Data}} \\ \cline{3-7}
\textbf{Data} & & BD-TTS & BD-VC & BD-NV & BD & BD-UT \\ \hline\hline 
BD-TTS & \multirow{3}{*}{AASIST} & \textbf{0.44} & \underline{16.85} & \underline{25.66} & \textbf{14.26} & \textbf{0.53} \\ 
BD-VC  & & \underline{18.71} & \textbf{2.18} & 35.31 & \underline{20.90} & \underline{14.34} \\ 
BD-NV  & & 23.44 & 41.63 & \textbf{9.53} & 26.30 & 26.87 \\ 
\hline
BD-TTS & \multirow{3}{*}{\makebox[1.3cm][c]{\parbox[c][2.5em][c]{1.5cm}{\centering W2V+ AASIST}}} & \textbf{1.01} & \underline{9.78} & \underline{14.34} & \textbf{8.08} & \textbf{0.20}  \\ 
BD-VC  & & \underline{5.81} & \textbf{3.82} & 18.26 & \underline{8.81} & \underline{9.35}\\ 
BD-NV  & & 9.34 & 17.38 & \textbf{7.77} & 11.33 & 23.79 \\ 
\Xhline{1px} 
\end{tabular}
}
\end{table}

\begin{table*}[bp]
\caption{Performance evaluation (EER\%) on test sets across various languages for models trained on English and Chinese at different epochs. ``9 langs” represents the combination of the 9 primary languages, while ``others” refers to the combination of remaining languages.}
\label{tab:lang}
\centering
\resizebox{\textwidth}{!}{
\begin{tabular}{c|c|p{1.1cm}<{\centering}|p{1.1cm}<{\centering}|p{1.1cm}<{\centering}|p{1.1cm}<{\centering}|p{1.1cm}<{\centering}|p{1.1cm}<{\centering}|p{1.1cm}<{\centering}|p{1.1cm}<{\centering}|p{1.1cm}<{\centering}|p{1.1cm}<{\centering}|p{1.1cm}<{\centering}}
\Xhline{1px} 
\multirow{2}{*}{\textbf{Model}} & \multirow{2}{*}{\textbf{Epoch}} & \multicolumn{11}{c}{\textbf{Test Data}}  \\ \cline{3-13}
 & & en & zh & es & fr & hi & ja & ko & fa & it & 9 langs & others \\ \hline\hline
\multirow{2}{*}{AASIST} & 20 & 0.81 & 2.14 & 14.60 & 22.54 & 26.06 & 9.53 & 4.39 & 9.73 & 6.79 & 10.86 & 6.03 \\ 
& \cellcolor{gray!15}50 & \cellcolor{gray!15}0.60 & \cellcolor{gray!15}3.48 & \cellcolor{gray!15}3.74 & \cellcolor{gray!15}9.70 & \cellcolor{gray!15}20.25 & \cellcolor{gray!15}8.95 & \cellcolor{gray!15}3.46 & \cellcolor{gray!15}8.62 & \cellcolor{gray!15}5.18 & \cellcolor{gray!15}8.49 & \cellcolor{gray!15}4.93 \\\hline
\multirow{2}{*}{W2V+AASIST} & 20 & 0.27 & 1.38 & 7.98 & 12.08 & 11.90 & 5.14 & 2.54 & 4.42 & 3.48 & 6.53 & 4.64 \\
&\cellcolor{gray!15} 50 & \cellcolor{gray!15}0.15 & \cellcolor{gray!15}0.29 & \cellcolor{gray!15}0.12 & \cellcolor{gray!15}0.42 & \cellcolor{gray!15}0.98 & \cellcolor{gray!15}0.22 & \cellcolor{gray!15}0.03 & \cellcolor{gray!15}0.24 & \cellcolor{gray!15}0.16 & \cellcolor{gray!15}0.50 & \cellcolor{gray!15}0.23 \\
\Xhline{1px} 
\end{tabular}
}
\end{table*}

For each training set, the best detection performance is consistently observed on its corresponding testing set, but performance degrades significantly when tested on other generator types. This highlights the challenge of generalizing across unseen generation methods.

In terms of overall performance, models trained on TTS data consistently deliver the best results on the full BD test set, followed by VC, while NV-trained models generally show lower performance. This is likely due to the TTS subset’s diverse composition, which includes state-of-the-art techniques that produce highly realistic synthetic speech. In contrast, NV-based systems may underperform because they often rely on older methods that generate lower-quality deepfakes, making detection more challenging for models trained on NV data.

When tested on the unseen commercial TTS API set (BD-UT), TTS-trained models consistently outperform those trained on VC and NV, achieving strong performance across both. This underscores that exposure to modern TTS data enhances the model’s ability to detect high-quality, natural-sounding deepfakes. 

In summary, unseen generation methods present a significant challenge for generalization in deepfake detection. Although training on similar generation types can somewhat improve detection performance, substantial differences between generation methods still result in considerable performance degradation. 
Additionally, we provide cross-evaluation results for individual latest generation methods in Figure~\ref{fig:eer} in the Appendix, which further confirm these findings.

\subsection{Cross-Lingual performance}
\label{subsec:language}
To assess the impact of language on deepfake detection, we conducted experiments using MD, where all generation methods were seen during training, but certain languages were kept unseen. The training set includes only English and Chinese, while the test set spans a total of 46 languages.

From Table~\ref{tab:lang}, we observe that both models perform well on the seen English (en) and Chinese (zh) test sets, with minimal error rates after just 20 epochs.
However, for the unseen languages, both models show a noticeable performance drop after 20 epochs, particularly for French (fr) and Hindi (hi). Extending the training to 50 epochs, AASIST still exhibits a significant gap between the seen and unseen languages, though there is some improvement for the unseen languages and minimal improvement for the seen ones. In contrast, the W2V+AASIST model achieves generally good performance, which can likely be attributed to the multilingual pretraining of the Wav2Vec 2.0 XLSR model~\cite{babu2021xlsr}.

These results suggest that language content does affect detection performance, even when the generation methods are seen during training. However, prior exposure to a language through multilingual pretraining can help mitigate this effect to some extent.

\subsection{Cross-Speaker Performance}
\label{subsec:speaker}
Some TTS systems are limited to generating specific voices, making it possible to detect deepfakes by merely memorizing the speaker’s voice rather than learning the distinct audio characteristics that differentiate real and fake speech. This raises the question: can a model learn to detect deepfakes based on their inherent characteristics, or does it simply overfit to the speaker identity?

To explore this, we created a small dataset selected from BD. To minimize the influence of different generation methods, we exclusively used TorToiSe~\cite{betker2023better}, a TTS system that supports multi-speaker speech generation. The training dataset is a subset of the BD train set, consisting of 100 real speakers and 10 fake speakers, with a total of 34,305 utterances. As detailed in Table~\ref{tab:speaker}, we designed five different test trials, varying the combinations of seen and unseen speakers to assess the model’s ability to generalize across speakers. For evaluation, we trained an AASIST model over three runs for 50 epochs on this training set.


\begin{table}[tb]
\centering
\caption{Statistics and EER(\%) results of cross-speaker testing trials. \#utt, \#spk represent number of utterances and speakers, respectively. The numbers in parentheses represent the distribution of speakers (seen, unseen) in the training set.}
\label{tab:speaker}
\resizebox{0.48\textwidth}{!}{
\begin{tabular}{c|cc|cc|c}
\Xhline{1px}
\multirow{2}{*}{\textbf{No.}} & \multicolumn{2}{c|}{\textbf{Real}} & \multicolumn{2}{c|}{\textbf{Fake}} & \multicolumn{1}{c}{\multirow{2}{*}{\textbf{EER(\%)}}} \\ \cline{2-5}
 & \textbf{\#utt} & \textbf{\#spk} & \textbf{\#utt} & \textbf{\#spk} &   \\ \hline\hline
1 & 6,599 & 100 (100, 0)  & 13,871 & 10 (10, 0) & 0.06 {\scriptsize $\pm$ 0.01}  \\
2 & 5,557 & 100 (0, 100) & 12,377 & 10 (0, 10) & 0.43 {\scriptsize$\pm$ 0.15} \\
3 & 5,557 & 100 (0, 100) & 13,871 & 10 (10, 0) & 0.01 {\scriptsize$\pm$ 0.01} \\
4 & 6,599 & 100 (100, 0) & 12,377 & 10 (0, 10) & 0.64 {\scriptsize$\pm$ 0.06} \\
5 & 6,071 & 100 (50, 50) & 13,677 & 10 (5, 5) & 0.49 {\scriptsize$\pm$ 0.05} \\
\Xhline{1px}
\end{tabular}
}
\end{table}

Overall, the EERs across all five test settings are minimal, indicating that the model can detect deepfake-specific features rather than relying solely on speaker identity. 
When comparing Settings 1 and 2, which differ in whether both real and fake speakers are seen or unseen during training, we observe only a slight increase in EER when speakers are unseen (from 0.06\% to 0.43\%).
In Setting 3, where real speakers are unseen and fake speakers are seen, the model achieves almost perfect detection (0.01\%), likely due to more fake data per speaker, though some speaker memorization may be occurring.
In contrast, Setting 4, with seen real speakers and unseen fake speakers, results in a higher EER (0.64\%), suggesting that the model struggles more with unseen fake speakers, possibly relying on learned fake speaker characteristics. Setting 5, with a mix of seen and unseen speakers, yields an EER of 0.49\%, indicating better generalization than Setting 4, but still some performance drop with unseen fake speakers.

The experimental results demonstrate that the model effectively learns deepfake-specific features instead of overfitting to individual speaker identities. While the impact of speaker identity on detection performance is generally minimal, it becomes more pronounced when the model encounters completely unseen fake speakers.

\section{Conclusion}
In conclusion, SpeechFake addresses critical gaps in existing speech deepfake detection datasets by providing a large-scale collection of over 3 million deepfakes, with diverse generation methods and languages. Through extensive experimentation, we established baseline results and demonstrated significant performance improvements for models trained on SpeechFake, particularly on unseen test sets. Our analysis of key factors, including generation methods, language diversity, and speaker variation, shows that while generation methods and language diversity influence detection performance, speaker variation has minimal impact. These findings highlight the challenges of generalizing across unseen deepfakes, while showcasing SpeechFake’s potential to advance model robustness and generalization. We believe SpeechFake will be an invaluable resource for developing robust detection systems, ultimately helping mitigate the risks of deepfake misuse.

\section{Limitations}
While SpeechFake provides a large and diverse dataset for speech deepfake detection, several limitations exist. First, although the dataset includes 40 different speech generation tools, it does not cover all current or emerging techniques. This is due to the rapid pace of advancements in speech generation technology, which introduces new methods that may not yet be represented. Additionally, the multilingual dataset is limited in terms of generation method variety, primarily because multilingual speech generation systems are still scarce. 
For future work, we plan to address these limitations by continuously updating the dataset to incorporate emerging generation techniques and expanding its multilingual component as new techniques become available.

\section{Acknowledgements}
This work was partially supported by the Ant Group Research Intern Program and the Pioneer R\&D Program of Zhejiang Province (No. 2024C01024). This work was supported in part by China NSFC projects under Grants 62122050 and 62071288, in part by Shanghai Municipal Science and Technology Commission Project under Grant 2021SHZDZX0102.


\appendix

\section{Dataset Details}
\subsection{Dataset Partition}
Table~\ref{tab:basic} provides the partition of SpeechFake as additional details for Section~\ref{sec:collect}.

\subsection{Dataset Metadata}
SpeechFake provides detailed metadata for each generated speech sample, including:
\begin{itemize}[leftmargin=*,itemsep=0pt]
\item Basic Labels: Identifying real or fake speech.
\item Generation Method: Specifying the tool used to create the speech.
\item Speaker/Voice ID: Providing identity labels for the original or generated voice.
\item Language ID: Indicating the language of the audio sample.
\item Text Transcriptions: Providing the corresponding text for the generated speech.
\end{itemize}

\subsection{Generation Methods}
Table~\ref{tab:list} presents a comprehensive list of the 40 speech generation tools used to create the SpeechFake dataset. These tools encompass a wide range of techniques, including TTS, VC, and NV systems. Notably, some methods, such as Fish Speech and CosyVoice, can be applied to multiple generation tasks (e.g., TTS and VC). For the 30 open-source tools, we carefully reviewed their licenses to ensure compliance with the construction and release of a publicly available dataset. The remaining 10 generation tools are commercial APIs, for which we obtained paid access, ensuring compliance with non-commercial research usage policies.

\subsection{License and Ethics}
The main part of the dataset is distributed under the CC BY 4.0 license, with certain portions licensed under alternative terms (e.g., GPL-3.0) to meet the requirements of specific tools used in dataset creation. 

To clarify, the dataset does not include deepfakes of identifiable real individuals. The voices in the dataset originate from either original training data (TTS), reference voices (VC), or original audio (NV), none of which are identifiable. Furthermore, all text and speech samples used in the dataset are sourced from publicly available speech datasets commonly used in speech generation research, and do not contain harmful or sensitive content.

\section{Experiment Details}
\subsection{Experimental Settings}
\label{app:exp}
Table~\ref{tab:train} outlines the training configurations for the two state-of-the-art models used in our experiments. The basic settings are consistent with the training setup proposed by \cite{tak2022automatic}. Unlike previous research on deepfake detection, we opted not to apply data augmentation in order to isolate the fundamental effects of the audio data and avoid potential biases introduced by augmentation methods, which may not generalize well across all datasets. Given the imbalance between deepfake and real samples, we employed weighted cross-entropy loss to ensure balanced training. Both models were trained for 50 epochs on 8 A100 GPUs over 1 run in the main experiments.  

\begin{table}[htb]
\centering
\caption{Training configurations of AASIST and W2V+AASIST models used in experiments.}
\label{tab:train}
\resizebox{0.48\textwidth}{!}{
\begin{tabular}{l|p{1.8cm}<{\centering}p{2.4cm}<{\centering}}
\Xhline{1px} 
\textbf{Configurations} & \textbf{AASIST} & \textbf{W2V+AASIST} \\ \hline
Model Size & 297K & 3M \\
Input Audio & \multicolumn{2}{c}{Chunk or pad to 4s}  \\
Data augmentation & None & None \\
Optimizer             & Adam & Adam       \\
Learning Rate         & 1e-4   & 1e-6       \\
Weight Decay          & 1e-4 & 1e-4       \\
Batch Size            & 1024      & 512          \\
Total Epochs          & 50 & 50          \\
\multirow{2}{*}{Loss Function}         & \multicolumn{2}{c}{Weighted Cross Entropy}  \\
& \multicolumn{2}{c}{(0.9 for real, 0.1 for fake)} \\
\Xhline{1px} 
\end{tabular}
}
\end{table}

For the test settings in Section~\ref{subsec:overall}, the following evaluation protocols were used:
\begin{itemize}[leftmargin=*,itemsep=0pt]
\item ASV19: original evaluation set.
\item FOR: randomly selected 10,000 utterances due to the small size of the original dataset.
\item WF: randomly selected 15,000 clips, as no predefined train/test splits exist.
\item ITW: entire dataset.
\item CDADD: original test set.
\item ASV24: development set, as evaluation labels were unavailable.
\end{itemize}

\subsection{Cross-model Evaluation}
\begin{figure}[t]
    \centering
    \includegraphics[width=0.43\textwidth]{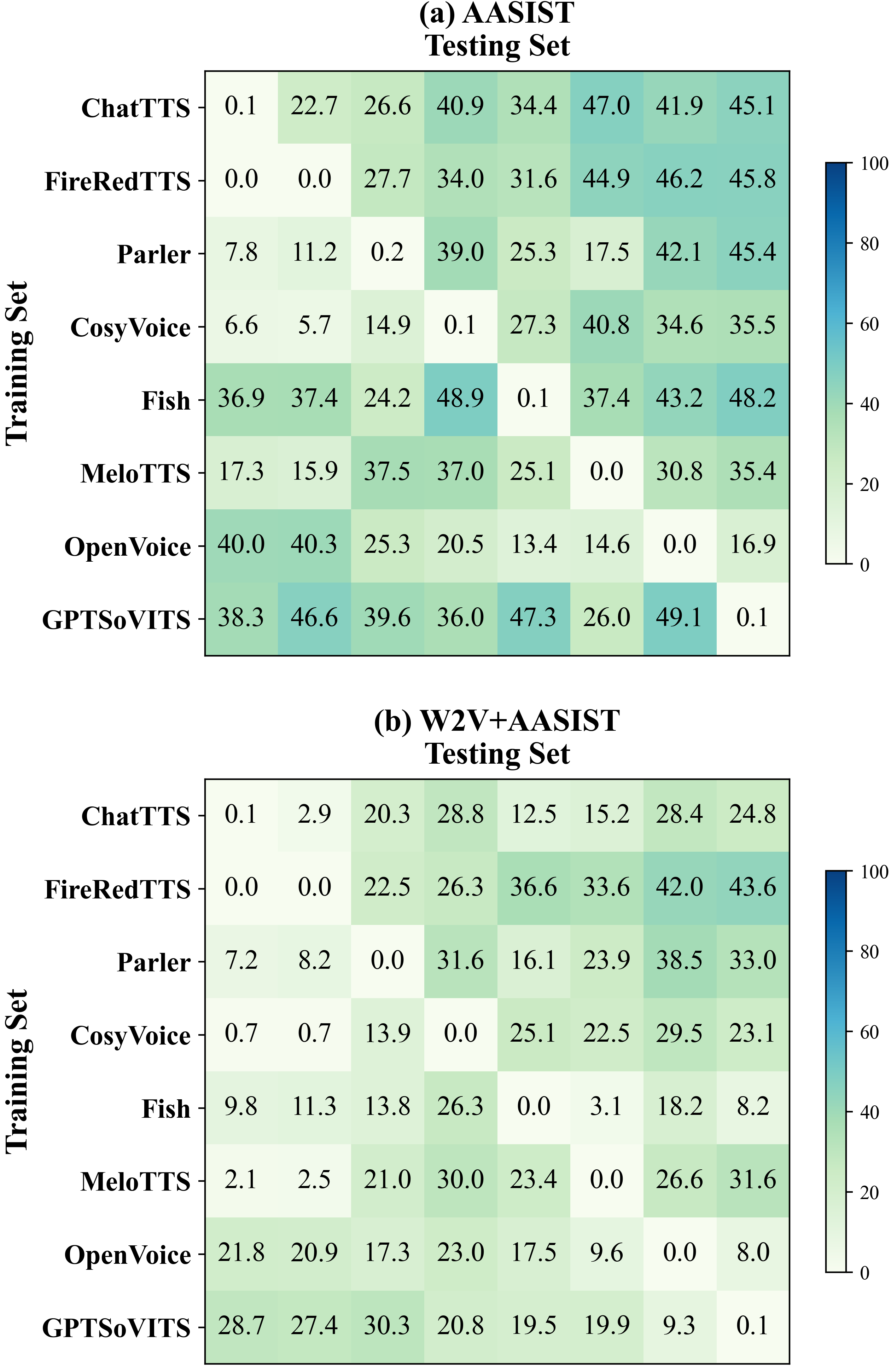}
    \caption{Cross-evaluation performance (EER\%) of models trained on subsets with individual generation methods, evaluated on their respective test sets. Eight latest generation methods were selected.}
    \label{fig:eer}
\end{figure}

Building on the cross-generator performance evaluation in Section~\ref{subsec:generator}, we assess the impact of the latest generation methods by evaluating models trained on individual ones. As shown in Figure~\ref{fig:eer}, the EER remains low on corresponding test sets but drops significantly on others. Some models perform well on specific unseen test sets (e.g., FireRedTTS-trained model on ChatTTS), but results are inconsistent across all sets. This highlights the challenge of generalizing to unseen deepfakes and the need for more robust detection models that can adapt to diverse generation methods.

\begin{table*}[htbp]
\centering
\caption{Partition of the SpeechFake dataset, with real and fake data divided into train, dev, test, and optional hidden sets.}
\resizebox{0.9\textwidth}{!}{
\begin{tabular}{c||ccc|c|ccc|c|c}
\Xhline{1px} 
\multirow{2}{*}{\textbf{Set}} & \multicolumn{4}{c|}{\textbf{Real Data}} & \multicolumn{5}{c}{\textbf{Fake Data}} \\ \cline{2-10}
& \textbf{train} & \textbf{dev} & \textbf{test} & \textbf{total} & \textbf{train} & \textbf{dev} & \textbf{test} & \textbf{hidden} & \textbf{total} \\ \hline\hline
BD & 75,708 & 12,618 & 37,854 & 126,180 & 633,354 & 105,544 & 315,222 & 898,896 & 1,953,016  \\ 
BD-UT & - & - & 37,854 & 37,854 & - & - & 50,000 & - & 50,000 \\
\hline
BD-EN & 38,400 & 6,400 & 19,200 & 64,000 & 389,866 & 64,970 & 193,461 & 480,585 & 1,128,882  \\ 
BD-CN & 37,308 & 6,218 & 18,654 & 62,180 & 243,488 & 40,575 & 121,760 & 418,311 & 824,134 \\ \hline
BD-TTS & 75,708 & 12,618 & 37,854 & 126,180 & 280,622 & 46,764 & 138,834 & 547,887 & 1,014,107 \\ 
BD-VC & 75,708 & 12,618 & 37,854 & 126,180 & 192,210 & 32,031 & 96,115 & 214,849 & 535,205 \\ 
BD-NV & 75,708 & 12,618 & 37,854 & 126,180 & 160,522 & 26,749 & 80,273 & 136,160 & 403,704 \\ \hline
MD & 60,000 & 10,000 & 152,757 & 222,757 & 208,126 & 34,690 & 726,136 & 366,540 & 1,335,492 \\
\Xhline{1px} 
\end{tabular}
}
\label{tab:basic}
\end{table*}

\begin{table*}[htbp]
\centering
\caption{List of generation methods used in the creation of SpeechFake.}
\label{tab:list}
\resizebox{\textwidth}{!}{
\begin{tabular}{clcl}
\Xhline{1px} 
\textbf{No.} & \textbf{Method} & \textbf{Generator} & \textbf{Link} \\ \hline
1 & MelGAN~\citep{kumar2019melgan} & NV & \url{https://github.com/kan-bayashi/ParallelWaveGAN} \\
2 & WaveGlow~\citep{prenger2019waveglow} & NV & \url{https://github.com/NVIDIA/waveglow} \\
3 & Parallel WaveGAN~\citep{yamamoto2020parallel} & NV & \url{https://github.com/kan-bayashi/ParallelWaveGAN} \\
4 & HiFi-GAN~\citep{kong2020hifi} & NV & \url{https://github.com/kan-bayashi/ParallelWaveGAN} \\
5 & Fullband-MelGAN~\citep{yang2021multi} & NV & \url{https://github.com/kan-bayashi/ParallelWaveGAN} \\
6 & StyleMelGAN~\citep{mustafa2021stylemelgan} & NV & \url{https://github.com/kan-bayashi/ParallelWaveGAN} \\
7 & FastDiff~\citep{huang2022fastdiff} & NV & \url{https://github.com/Rongjiehuang/FastDiff} \\
8 & BigVGAN~\citep{lee2023bigvgan} & NV & \url{https://github.com/NVIDIA/BigVGAN} \\
9 & WaveNet~\citep{van2016wavenet} & TTS  & \url{https://github.com/r9y9/wavenet_vocoder} \\
10 & Tactotron2~\citep{shen2018natural} & TTS & \url{https://github.com/NVIDIA/tacotron2} \\
11 & Glow-TTS~\citep{kim2020glow} & TTS & \url{https://github.com/jaywalnut310/glow-tts} \\
12 & Grad-TTS~\citep{popov2021grad} & TTS & \url{https://github.com/huawei-noah/Speech-Backbones} \\
13 & FastSpeech2~\citep{ren2021fastspeech} & TTS  & \url{https://github.com/ming024/FastSpeech2} \\
14 & PortaSpeech~\citep{ren2021portaspeech} & TTS & \url{https://github.com/keonlee9420/PortaSpeech} \\
15 & VITS~\citep{kim2021conditional} & TTS & \url{https://github.com/jaywalnut310/vits} \\
16 & StarGAN-VC~\citep{li2021stargan} & VC & \url{https://github.com/yl4579/StarGANv2-VC} \\
17 & DiffGAN-TTS~\citep{liu2022diffgan} & TTS & \url{https://github.com/keonlee9420/DiffGAN-TTS} \\
18 & ProDiff-TTS~\citep{huang2022prodiff} & TTS & \url{https://github.com/Rongjiehuang/ProDiff} \\
19 & EdgeTTS & TTS & \url{https://github.com/rany2/edge-tts.git} \\
20 & TorToiSe~\citep{betker2023better} & TTS & \url{https://github.com/neonbjb/tortoise-tts} \\
21 & StyleTTS2~\citep{li2024styletts} & TTS & \url{https://github.com/yl4579/StyleTTS2} \\
22 & OpenVoice~\citep{qin2023openvoice} & VC & \url{https://github.com/myshell-ai/OpenVoice} \\
23 & GPTSoVITS & VC & \url{https://github.com/RVC-Boss/GPT-SoVITS} \\
24 & Fish Speech~\citep{fish-speech-v1.4} & TTS/VC & \url{https://github.com/fishaudio/fish-speech} \\
25 & MeloTTS & TTS & \url{https://github.com/myshell-ai/MeloTTS} \\
26 & ChatTTS & TTS & \url{https://github.com/2noise/ChatTTS} \\
27 & CosyVoice~\citep{du2024cosyvoice} & TTS/VC & \url{https://github.com/FunAudioLLM/CosyVoice} \\
28 & Parler-TTS~\citep{lyth2024natural} & TTS & \url{https://github.com/huggingface/parler-tts} \\
29 & FireRedTTS~\citep{guo2024fireredtts} & TTS & \url{https://github.com/FireRedTeam/FireRedTTS} \\
30 & Seed-VC~\citep{liu2024zero} & VC & \url{https://github.com/Plachtaa/seed-vc} \\
\midrule
31 & Volcengine API & TTS & \url{https://www.volcengine.com} \\
32 & Baidu API & TTS & \url{https://cloud.baidu.com} \\
33 & AliYun API & TTS & \url{https://www.aliyun.com} \\
34 & Xfyun API & TTS & \url{https://www.xfyun.cn} \\
35 & Moyin API & TTS & \url{https://www.moyin.com} \\
36 & Microsoft API & TTS & \url{https://azure.microsoft.com} \\
37 & Google API & TTS & \url{https://cloud.google.com} \\
38 & Amazon API & TTS & \url{https://docs.aws.amazon.com/polly} \\
39 & OpenAI API & TTS & \url{https://platform.openai.com} \\
40 & GPT4o API & TTS & \url{https://platform.openai.com} \\

\Xhline{1px} 
\end{tabular}
}
\end{table*}

\end{document}